# Nuclear magnetic resonance implementation of the Deutsch-Jozsa algorithm using different initial states

**Vladimir L. Ermakov[1], and B. M. Fung[2]**

*Department of Chemistry and Biochemistry, University of Oklahoma,*

*Norman, Oklahoma 73019-3051, USA*

[1]On leave from Kazan Physical-Technical Institute, Kazan 420029, Russia.

E-mail: ermakov@ou.edu (current); ermakov@smtp.ru (permanent).

[2]To whom correspondence should be addressed. E-mail: bmfung@ou.edu.

## Abstract

The Deutsch-Jozsa algorithm distinguishes constant functions from balanced functions with a single evaluation. In the first part of this work, we present simulations of the nuclear magnetic resonance (NMR) application of the Deutsch-Jozsa algorithm to a 3-spin system for all possible balanced functions. Three different kinds of initial states are considered: a thermal state, a pseudopure state, and a pair (difference) of pseudopure states. Then, simulations of several balanced functions and the two constant functions of a 5-spin system are described. Finally, corresponding experimental spectra obtained by using a 16-frequency pulse to create an input equivalent to either a constant function or a balanced function are presented, and the results are compared with those obtained from computer simulations.

## I. INTRODUCTION

The earliest quantum algorithm that demonstrates the advantage of quantum algorithms over classical ones is the Deutsch-Jozsa algorithm,[1,2] which deals with the analysis of binary functions. A function that transforms *n*-bits of binary information to one bit

$$f(x): \quad \{0, 1\}^n \mapsto \{0, 1\} \qquad (1)$$

is either a constant function or a balanced function. A constant function has an output of $f(x) = 0$ or $f(x) = 1$ for all $x$; a balanced function has $f(x) = 0$ for half of the inputs and $f(x) = 1$ for the other half. To determine whether a function is constant or balanced, a classical algorithm requires up to $(2^{n-1} + 1)$ evaluations, but the Deutsch-Jozsa quantum algorithm requires a single evaluation regardless of the input size, *n*, by mapping *n* qubits to one qubit. In a modification called the refined Deutsch-Jozsa algorithm,[3] one fewer qubit is required.

The Deutsch-Jozsa algorithm has been implemented by NMR using different approaches.[4-14] In its original version, the algorithm requires a pure initial state. A thermal state can be considered as a weighted sum over all pure states, and most quantum algorithms cannot be implemented by starting with a thermal state. However, experimental nuclear magnetic resonance (NMR) realizations show that the algorithm also works with a thermal state as the starting state.[4,6,13] In other words, the output spectra of a spin system display different patterns for constant functions and balanced functions even when the initial state is at thermal equilibrium rather than being a pseudopure state. A general procedure for solving the Deutsch-Jozsa problem adapted to the thermal state has been described.[14]

For the implementation of the Deutsch-Jozsa algorithm, the only comparison between the results obtained using a pseudopure state and a thermal state has been reported for a simple 2-spin system,[4] but such a comparison should be made for larger spin systems for the benefit of further development in this subject. Another practical aspect to be considered is the experimental complexity in the preparation of pseudopure spin states. For example, using temporal averaging to prepare a pseudopure state for a 5-spin system, 124 NMR experiments would be required;

therefore, only part of a pseudopure state was prepared to examine the Deutsch-Jozsa algorithm.[7] As an alternative, the preparation of the algebraic sum of a pair (difference) of pseudopure states (POPS) by using a very simple method has been proposed,[15,16] but this approach has not been applied to implement a bona fide quantum algorithm. It has also been suggested that mixed (sum) superposition states can be used for certain computational problems,[17] and experimental results for logic operations of a single qubit in a 3-qubit system were given.

In the NMR implementation of quantum algorithms, the unitary transform is usually carried out by using a series of pulses with spacing related to the coupling constants. Under certain circumstances, the elaborate pulse sequences can be replaced by simpler ones containing shaped multi-frequency pulses designed to excite a number of transitions simultaneously.[6,10,18] These multi-frequency pulses are also useful for the preparation of pseudopure states[19] and for quantum simulations.[20] The aim of this work is two-fold: to demonstrate how the Deutsch-Josza algorithm can be implemented to a 5-spin system by using a simple pulse sequence with moderately complex multi-frequency pulses, and to demonstrate that the POPS approach is indeed a viable alternative for the implementation of quantum algorithms.

The paper is organized as follows. First, the NMR version of the algorithm used in this work, which follows the approach of Linden *et al.*,[6] is briefly described. Then, the results of computer simulations of a 3-spin system starting with three different initial states − a thermal state, a pseudopure state, and a POPS − are presented. Following this, computer simulations of a 5-spin system are shown. Finally, the experimental implementation of the Deutsch-Jozsa algorithm for the 5-spin system, as guided by the simulations, is described.

## II. NMR VERSION OF THE DEUTSCH-JOZSA ALGORITHM

There are two methods in applying NMR to implement the Deutsch-Jozsa algorithm. The first is the use of a series of spin-selective $\pi/2$ pulses separated by appropriate evolution times which are determined by the coupling constants between different spins.[4,5,7-13] Another approach is the use of transition-selective π-pulses to one of the spins (work spin) to realize the two types of functions.[6,10] Here we employ the second approach. The schematic diagram for the pulse sequence is shown in Fig. 1, using a three-spin system as an example. Since the details of the method have been described in the literature,[6] only a brief description is given here.

First, a hard (non selective) π/2 pulse is applied to all spins to produce a (pseudo) Hadamard transform, which is necessary to prepare the system in the uniform superposition state. Second, a shaped multi-frequency selective π-pulse is applied to the work spin to invert some of its peaks to mimic the functional form. For the 3-spin system illustrated, this corresponds to manipulating the four-peak spectrum on the left side of the figure; for a 5-spin system, a more complex multi-frequency pulse is needed for this step (16 frequencies for one of the constant functions and 8 for the balanced functions). The number and the order of the nonzero harmonics of this pulse define which function is realized: if all harmonics have the same amplitude (zero or nonzero), it is a constant function; if half of them have nonzero amplitude and the other half have zero amplitude, it is a balanced function. Finally, a free induction decay (FID) is recorded immediately after the selective pulse. It should be noted that the original quantum mechanical algorithm requires an inverse Hadamard transform as the final step; interestingly, in NMR implementation this step is simply realized by not applying a π/2 reading pulse – thus, no pulse is needed after the selective multi-frequency π-pulse.

## III. SIMULATIONS OF A 3-SPIN SYSTEM

For an *n*-bit system, there are two constant functions, and $C_{N/2}^{N}$ balanced functions, where $N = 2^n$ is the total number of system states. For example, for a three-spin NMR system, one spin can be used as a work qubit and two as data qubits, so that the number of constant functions is $C_2^4 = 6$. Similarly, there are $C_4^8 = 70$ balanced functions for a four-spin system (one work qubit and three data qubits), and $C_8^{16} = 12870$ balanced functions for a five-spin system (one work qubit and four data qubits). Thus, the Deutsch-Jozsa algorithm can be readily demonstrated, both experimentally[6,10] and by simulations, to examine all the balanced functions of a 3-spin system, but this is not practical for systems having five or more spins. Therefore, one must choose some of the balanced functions to demonstrate the experimental implementation of the algorithm for large spin systems. We use computer simulations to help the choice.

To examine all possible cases using different kinds of initial states, computer simulations are first carried out on a 3-spin system. This enables us to make comparisons and set guidelines for larger spin systems.

For the simulations, we consider three coupled spins A, B and C in an anisotropic medium with chemical shifts $\omega_A/2\pi = -20000$ Hz, $\omega_B/2\pi = 0$ Hz, and $\omega_C/2\pi = +15000$ Hz. Couplings between the spins are characterized by $\Delta = 2D + J$, where $D$ and $J$ are dipolar and scalar coupling constants, respectively: $\Delta_{AB} = 6000$ Hz, $\Delta_{AC} = 2000$ Hz, and $\Delta_{BC} = 500$ Hz. For spins in isotropic media ($D = 0$, usually $J < 200$ Hz, and possibly smaller chemical shifts), the results are qualitatively the same. The pulse sequence shown in Fig. 1 is applied to three different initial states: the thermal state $A_z + B_z + C_z$, the pseudopure state $|000\rangle\langle000|$, and a pair of pseudopure states $|000\rangle\langle000| - |100\rangle\langle100|$. Spin A is chosen as the work spin. Phased spectra are presented in the simulations because they are important for the comparison of the three cases.

*Thermal initial state*

The results for starting with the thermal state are shown in Fig. 2. The functions themselves are shown on the right side using parentheses. Each one of the four sequential positions inside the parentheses denotes an output (0 or 1) of a binary function when its input is equal to two-bit inputs: 00, 01, 10, or 11. The phases of the peaks of the work spin A indicate which function is applied in the algorithm (0 = positive and 1 = negative). In the spectra produced by applying the two constant functions (**a** and **b**), the peaks of the data spins B and C are the same as those of "doing nothing." In the spectra produced by the six balanced functions (**c** to **h**), the peaks of one or both data spins disappear. Thus, the two types of functions can be readily distinguished.

*Pseudopure initial state*

The results of simulations using the pseudopure initial state |000><000| are shown in Fig. 3. In this case, the spectra for the data spins B and C show different phase characteristics for the two types of functions: all 8 peaks have a positive phase for a constant function, but some peaks have a negative phase for a balanced function. When the absolute-value mode is used, all 8 functions would give the same spectra. In other words, the initial pseudopure state discriminates the functions only in the phase-sensitive mode.

Application of the algorithm to the other 7 pseudopure states (not shown in the figure) produces similar simulated spectra – the constant and balanced functions only differ in the phases of the peaks.

*A pair of pseudopure states as the initial state*

The results of simulations using a pair of pseudopure states (POPS), in this case the difference |000><000| − |100><100|, as the initial state are displayed in Fig. 4. This initial state produces spectra which allow the formulation of a very simple criterion to distinguish between the two types of functions: if no peaks are observed for the data spins, it is a constant function; otherwise (namely there are observable peaks for the data spins), it is a balanced function. The initial POPS that have such an effect are: |000><000| − |100><100|, |001><001| − |101><101|, |010><010| − |110><110|, and |011><011| − |111><111|. Each of these POPS is a difference of pseudopure states which differ only in the state of the work spin A, and each one can be prepared by selectively inverting one of the four peaks of this work spin.[15,16] A similar strategy is used to create POPS in the subsequent simulations of a 5-spin system and in experimental executions.

It must be pointed out that the use of POPS as the initial state has the distinct advantage that it can be employed for the absolute-value mode. Furthermore, unlike the initial thermal state, the spectra for constant functions are different from the "doing nothing" spectrum.

**IV. SIMULATIONS OF A 5-SPIN SYSTEM**

The 5-spin system used in this work is 2,4,5-trifluorobenzonitrile dissolved in a liquid crystalline solvent, which was utilized to demonstrate the preparation of pairs of pseudopure states (POPS).[16] All 16 peaks for each of the five spins are well resolved. In the first step of the study of this 5-spin system, computer simulations are made to help design actual experiments to realize the Deutsch-Jozsa algorithm for this 5-qubit system. In our notation, the protons are designated as spins A and B, and the fluorine nuclei are designated as spins C, D, and E. In the simulations the frequency difference between these two types of nuclei is reduced to allow both $^1$H and $^{19}$F spectra be displayed in a single figure. The chemical shifts used in the simulations are: $\omega_A/2\pi = 9770$ Hz, $\omega_B/2\pi = 9647$ Hz, $\omega_C/2\pi = -2961$ Hz, $\omega_D/2\pi = -7815$ Hz, and $\omega_E/2\pi = -13082$ Hz. The values of $\Delta = 2D + J$ are the same as the experimental values.[15] The pulse sequence shown in Fig. 1 is used.

In a previous study of 2-spin and 3-spin systems starting with a single pseudopure state, a special phase-cycling scheme was used so that the two types of functions could be distinguished in either the phased mode or the absolute-value mode of display: spectra for constant functions show peaks for the data spins, whereas balanced functions produce no peaks.[10] The method used is actually a modification of the Deutsch-Jozsa algorithm, but it was not stated explicitly. Experimentally, there is usually some arbitrariness in "phasing" a spectrum after the Fourier transform, which would cause substantial uncertainty in judgment for larger spin systems. On the other hand, there is no such problem when absolute-value spectra are studied.[6] However, the results presented in the previous section show that applying either a constant function or a balanced function to a single pseudopure state yields the same number of peaks which differ only in their phases; consequently, the absolute-value spectra would not be useful. Furthermore, because the preparation of a pseudopure state for a 5-spin system requires many elaborate steps, we did not investigate the use of a single pseudopure state as the initial state experimentally. Therefore, no corresponding simulation was carried out.

*Thermal initial state*

The simulated spectra (absolute-value mode) starting with a thermal initial state are displayed in Fig. 5. Following the previous notation, the functions themselves are shown on the right side using parentheses. Each of the 16 sequential numbers denotes an output (0 or 1) of a binary function with a 4-bit input: 0000, 0001, …., 1111. In this simulation and the subsequent experiment, the $^{19}$F spin "D" is chosen as the work spin. The spectra obtained from applying the two constant functions are shown in rows **a** and **b** in Fig. 5, and they are identical to the equilibrium "doing nothing" spectrum observed experimentally,[16] except that the small splittings are obscured in the simulated spectra. When any of the $C_8^{16}$ = 12870 balanced functions is applied, the spectrum shows fewer peaks. For the sake of clarity, we choose to display the simulated spectra of four balanced functions which result in complete destruction of the peaks for one of the four data spins (spectra **c**, **d**, **e**, and **f** in Fig. 5).

*A pair of pseudo pure state as initial state*

In section III it was shown that using a single pseudopure state as the starting state does not afford any distinction between constant and balanced functions in the absolute-value mode. On the other hand, the use of pairs of pseudopure states (POPS) is very effective. Besides, POPS has the advantage of being very simple to prepare experimentally.[15,16] For a 5-qubit system, there are 32 pseudopure states and 496 possible pairs (there is a mistake in these numbers in ref. 15), of which 80 POPS are experimentally accessible, but the results of only one POPS is demonstrated. Denoting the spin states in the order |ABCDE>, the POPS chosen is |00000><00000| − |00010><00010|, where the state of the work spin D is changed. Any other 15 POPS involving the change of the spin state of D would be equally effective. The pulse sequence is slightly more complicated than that shown in Fig. 1, and will be discussed in the next section.

The simulated spectra obtained by using the chosen POPS as the initial state are shown in Fig. 6. In this case, spectra for the constant functions exhibit no peak for the data spins (**a** and **b**), and are quite different from the "doing nothing" spectrum. For the balanced functions considered above, the peaks of only one data spin survive (**c** to **f**); this group of peaks corresponds to the group that is eliminated when the initial state is the thermal state.

## V. EXPERIMENTAL RESULTS AND DISCUSSION

All experiments were carried out on a Varian INOVA-400 NMR spectrometer at 21 °C. The sample was 5% 2,4,5-trifluorobenzonitrile dissolved in a liquid crystal mixture.[16]

When the thermal state is used as the initial state, the pulse sequence shown in Fig. 1 is directly applied. It can be represented as:

$$I_z - [\text{non-selective } \pi/2 \text{ pulse}] - [\text{multi-frequency selective } \pi \text{ pulse}] - FID \quad (2)$$

When a POPS is used as the initial state, a second experiment is needed. This experiment adds a single-line selective pulse that inverts one of the peaks in the work spin:

$$I_z - [\text{single-frequency selective } \pi \text{ pulse}] - [\text{non-selective } \pi/2 \text{ pulse}]$$
$$- [\text{multi-frequency selective } \pi \text{ pulse}] - FID \quad (3)$$

Then the FID obtained by applying (3) is subtracted from that obtained by applying (2), and the result is Fourier-transformed to obtain the spectrum. As remarked previously, only absolute-value spectra are displayed.

For experimental implementation of these sequences, we applied 16-frequency pulses corresponding to the two constant functions (**a** and **b** shown in Figs. 5 and 6) and two of the balanced functions (**c** and **d** shown in Figs. 5 and 6). The amplitudes of the harmonics were set to be proportional to the 0's and 1's in the corresponding functions depicted on the right hand side of the figures. Since the simulated $^1$H spectra for these four cases are identical for either type of initial state, only the $^{19}$F spectra were studied.

The results of applying sequence (2) to the thermal initial state $I_z$ are shown in Fig. 7. There is good agreement between the results of the experiments and simulations (Fig. 5). Although the peaks which are destroyed in the simulations are not completely suppressed experimentally due to pulse imperfection and quantum decoherence, the distinction between constant and balanced functions is clear: the two constant functions give spectra similar to "doing nothing"; for the two balanced functions studied, all peaks for one of the data spins have greatly reduced intensities. For other balanced functions, with the exception of the two functions depicted in the simulations in **e** and **f** of Figs. 5 and 6, peaks with reduced intensities would be distributed among two or more of the four data spins.

The results of using a POPS as the initial state are shown in Fig. 8. Again, the experimental spectra agree well with the simulated spectra (Fig. 6), except that the intensities of some of the peaks are reduced greatly instead of disappearing completely. The spectral characteristics can be summarized in the following: if the peaks for all the data spins are greatly diminished, the input is a constant function; otherwise, it is a balanced function. It should be noted that, unlike the case of using the thermal state as the initial state (Fig. 7), the spectra obtained for the constant functions are now completely different from the "doing nothing" spectrum.

In summary, we have shown that the Deutsch-Jozsa algorithm can be implemented on multi-spin systems by using a very simple pulse sequence with multi-frequency pulses. The use of POPS instead of a single pseudopure state as the initial state can readily distinguish the constant functions from balanced functions in the absolute-value mode without modifying the algorithm. It is conceivable that the POPS approach can also be used to implement other quantum algorithms.

ACKNOWLEDGEMENT

This work was supported by the National Science Foundation under grant number DMR-0090218.

# References

[1] D. Deutsch and R. Jozsa, Proc. R. Soc. London, Ser. A **439**, 553 (1992).

[2] R. Cleve, A. Ekert, C. Macchiavello, and M. Mosca, Proc. R. Soc. London, Ser. A **454**, 339 (1998).

[3] D. Collins, K. W. Kim and W. C. Holton, Phys. Rev. A **58**, R1633 (1998).

[4] I. L. Chuang, L. M. K. Vandersypen, X. Zhou, D. W. Leung, and S. Lloyd, Nature (London) **393**, 143 (1998).

[5] J. A. Jones and M. Mosca, J. Chem. Phys. **109**, 1648 (1998).

[6] N. Linden, H. Barjat, and R. Freeman, Chem. Phys. Lett. **296**, 61 (1998).

[7] R. Marx, A. F. Fahmy, J. M. Myers, W. Bermel, and S. J. Glaser, Phys. Rev. A **62**, 012310 (2000).

[8] D. Collins, K.W. Kim, W. C. Holton, H. Sierzputowska-Gracz, and E. O. Stejskal, Phys. Rev. A **62**, 022304 (2000).

[9] J. Kim, J.-S. Lee, S. Lee, and C. Cheong, Phys. Rev. A **62**, 022312 (2000).

[10] K. Dorai, Arvind, and A. Kumar, Phys. Rev. A **61**, 042306 (2000).

[11] K. Dorai, Arvind, and A. Kumar, Phys. Rev. A **63**, 034101 (2001).

[12] D. R. Sypher, I. M. Brereton, H. M. Wiseman, B. L. Hollis, and B. C. Travaglione, Phys. Rev. A **66**, 012306 (2002).

[13] D. Wei, J. Luo, X. Yang, X. Sun, X. Zeng, M. Liu, S. Ding, and M. Zhan, Prep. Arch., quant-ph/0301041 (2003).

[14] J. M. Myers, A. F. Fahmy, S. J. Glaser, and R. Marx, Phys. Rev. A **63**, 032302 (2001).

[15] B. M. Fung, Phys. Rev. A **63**, 22304 (2001).

[16] B. M. Fung, J. Chem. Phys. **115**, 8044 (2001).

[17] Z. L. Mádi, R. Brüschweiler, R. R. and Ernst, J. Chem. Phys. **109**, 10603 (1998).

[18] V. L. Ermakov and B. M. Fung, Phys. Rev. A **66**, 042310 (2002).

[19] A. K. Khitrin, H. Sun, and B. M. Fung, Phys. Rev. A **63**, 020301 (2001).

[20] A. K. Khitrin and B. M. Fung, Phys. Rev. A **64**, 032306 (2001).

**FIGURE CAPTIONS**

FIG. 1. The pulse sequence used in this work for NMR implementation of the Deutsch-Jozsa algorithm.

FIG. 2. The phase-sensitive simulated spectra of a 3-spin system. The application of the Deutsch-Jozsa algorithm to the initial thermal state is shown for all all eight functions. **a** and **b**: Constant functions; **c** to **f**: Balanced functions. Spin A is the work spin, and their phases indicate which function was applied in the algorithm. The functions are shown on the right side in parentheses.

FIG. 3. The phase-sensitive simulated spectra of a 3-spin system. The application of the Deutsch-Jozsa algorithm to the pseudopure initial state |000><000| is shown for all eight functions. **a** and **b**: Constant functions; **c** to **f**: Balanced functions. Spin A is the work spin, and their phases indicate which function was applied in the algorithm. The functions are shown on the right side in parentheses.

FIG. 4. The phase-sensitive simulated spectra of applying the Deutsch-Jozsa algorithm to a 3-spin system, using a pair of pseudopure states |000><000| − |100><100| as the initial state. **a** and **b**: Constant functions; **c** to **f**: Balanced functions. Spin A is the work spin, and their phases indicate which function was applied in the algorithm. The functions are shown on the right side in parentheses.

FIG. 5. The simulated spectra (absolute-value mode) of a 5-spin system. The application the Deutsch-Jozsa algorithm to the initial thermal state is shown for six functions. **a** and **b**: Constant functions; **c** to **f**: Balanced functions. Spin D is the work spin, and the functions are shown on the right side in parentheses. The small splittings are obscured in these spectra.

FIG. 6. The simulated spectra (absolute-value mode) obtained by applying the Deutsch-Jozsa algorithm to a 5-spin system, using a pair of pseudopure states (|00000><00000| − |00010><00010|) as the initial state. **a** and **b**: Constant functions; **c** and **d**: Balanced functions. Spin D is the work spin, and the functions are shown on the right side in parentheses. The small splittings are obscured in these spectra.

FIG. 7. Experimental $^{19}$F spectra (absolute-value mode) obtained by applying the Deutsch-Jozsa algorithm to a 5-spin system with an initial thermal state. **a** and **b**: Constant functions; **c** and **d**: Balanced functions. Spin D is the work spin; the functions are the same as those in Figs. 5 and 6. The small splittings are obscured in these spectra.

FIG. 8. Experimental $^{19}$F spectra (absolute-value mode) obtained by applying the Deutsch-Jozsa algorithm to a 5-spin system, using a pair of pseudopure states (|00000><00000| − |00010><00010|) as the initial state. **a** and **b**: Constant functions; **c** and **d**: Balanced functions. Spin D is the work spin; the functions are the same as those in Figs. 5 and 6. The small splittings are obscured in these spectra.

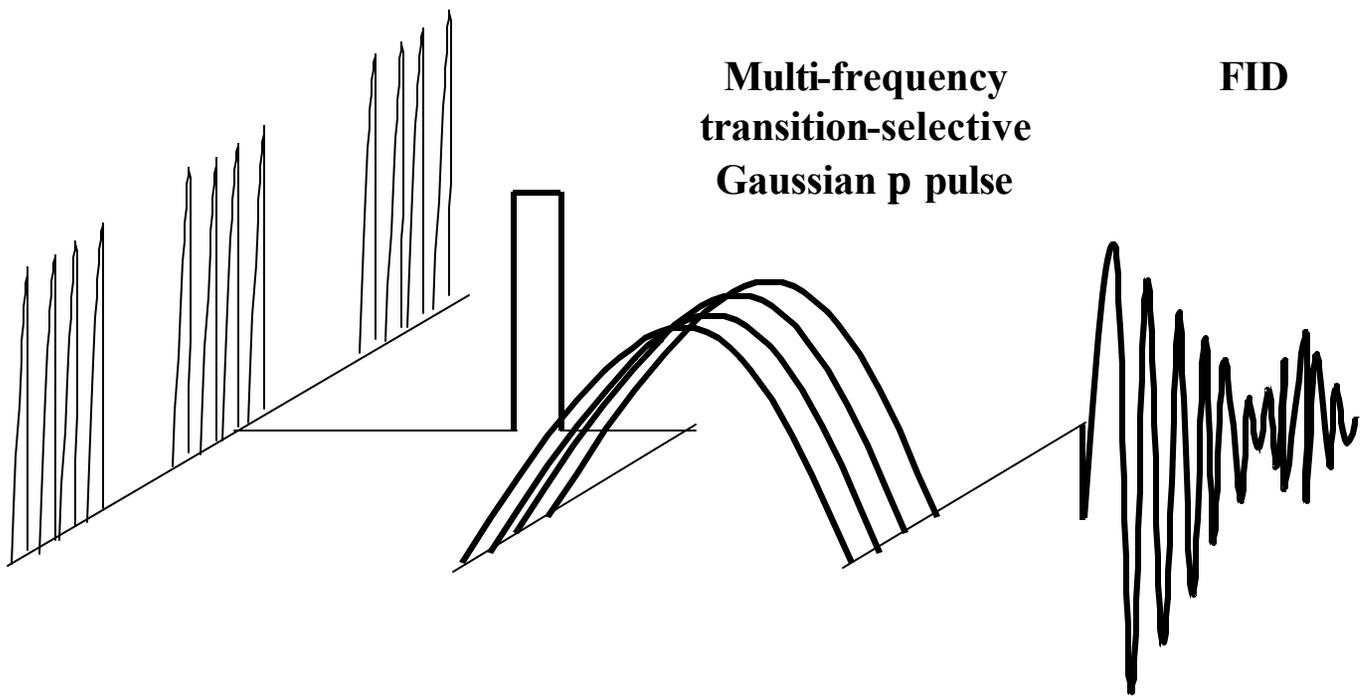

Fig. 1. V.L. Ermakov and B.M. Fung

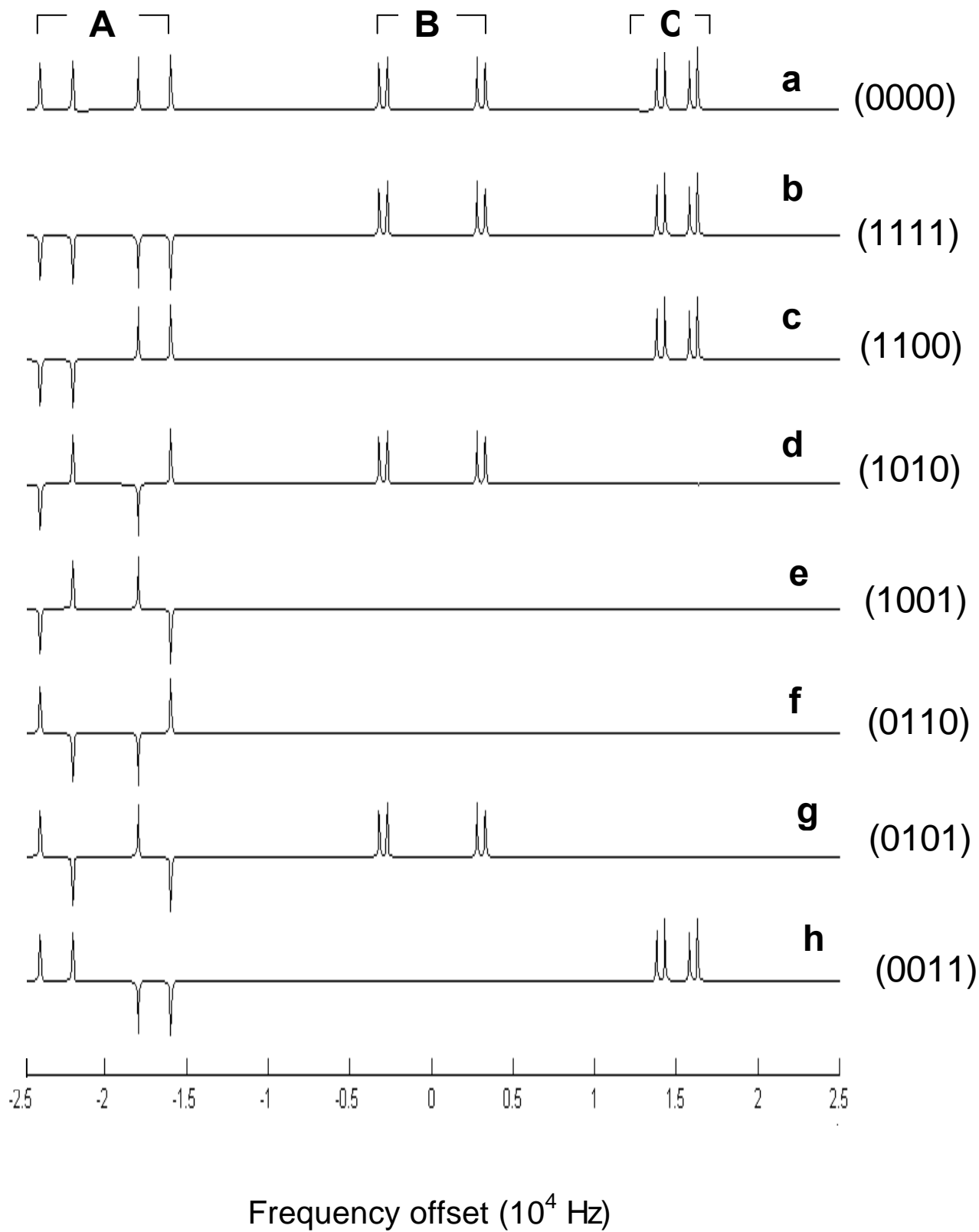

Fig. 2, V.L. Ermakov and B.M. Fung

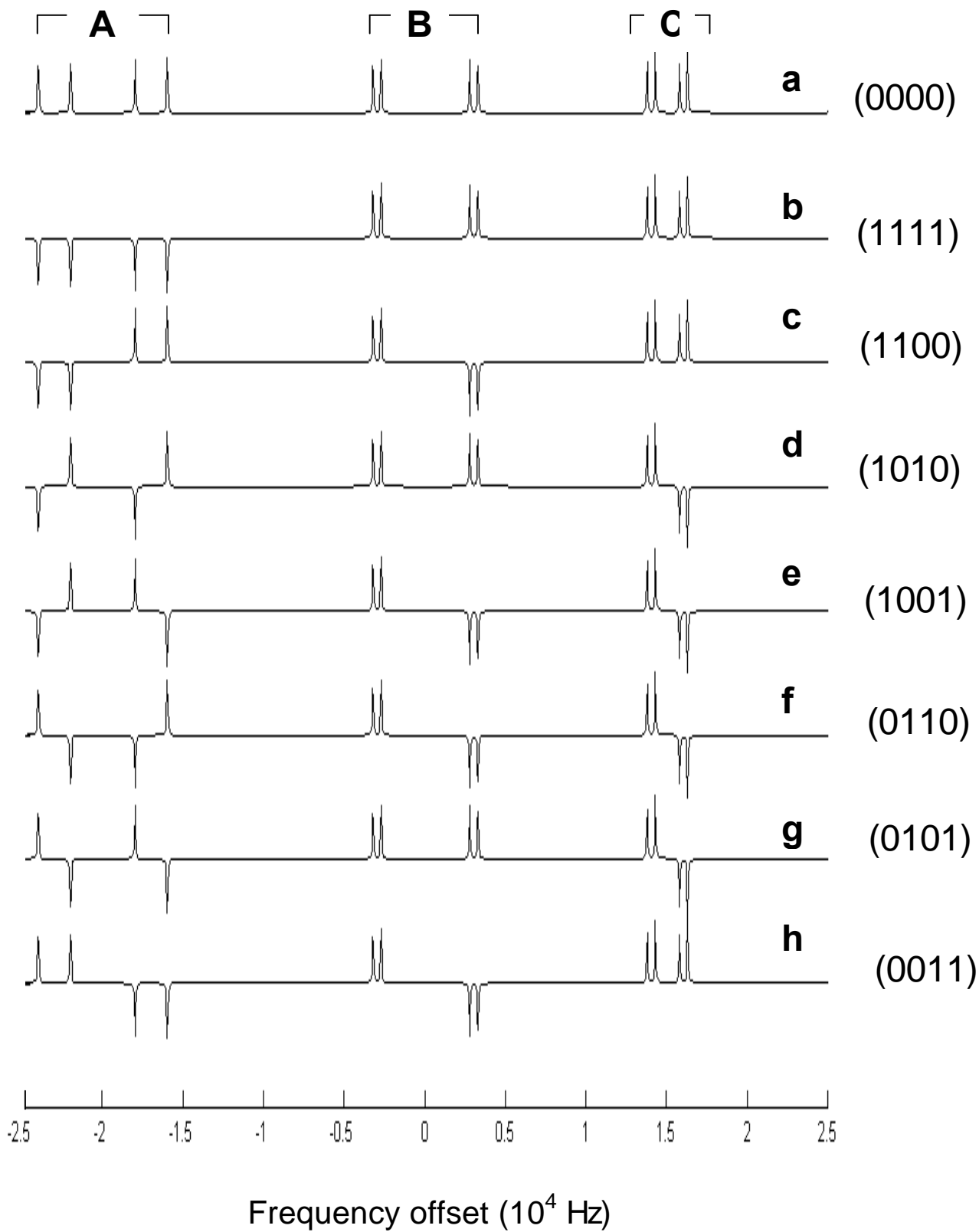

Fig. 3, V.L. Ermakov and B.M. Fung

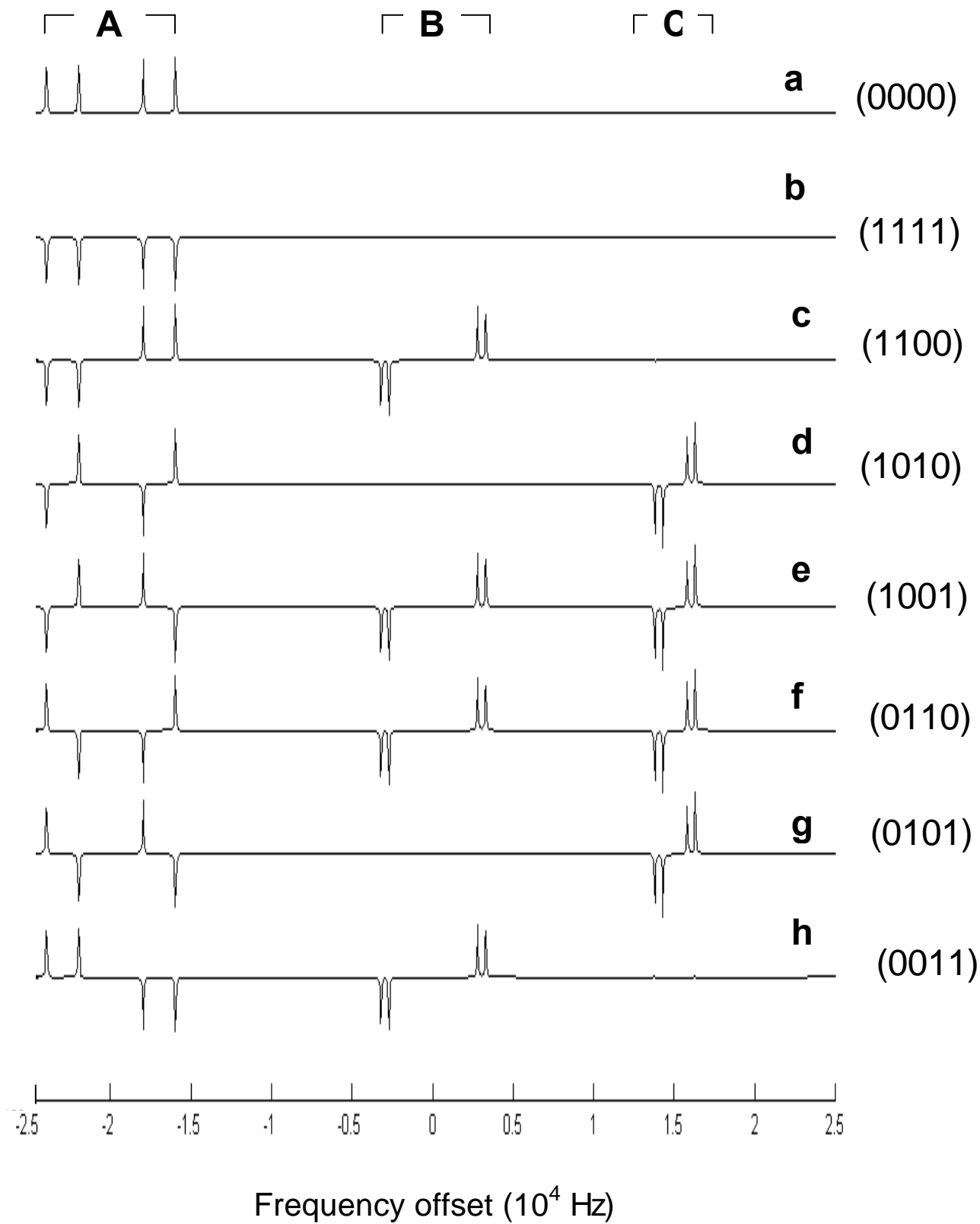

Fig. 4, V.L. Ermakov and B.M. Fung

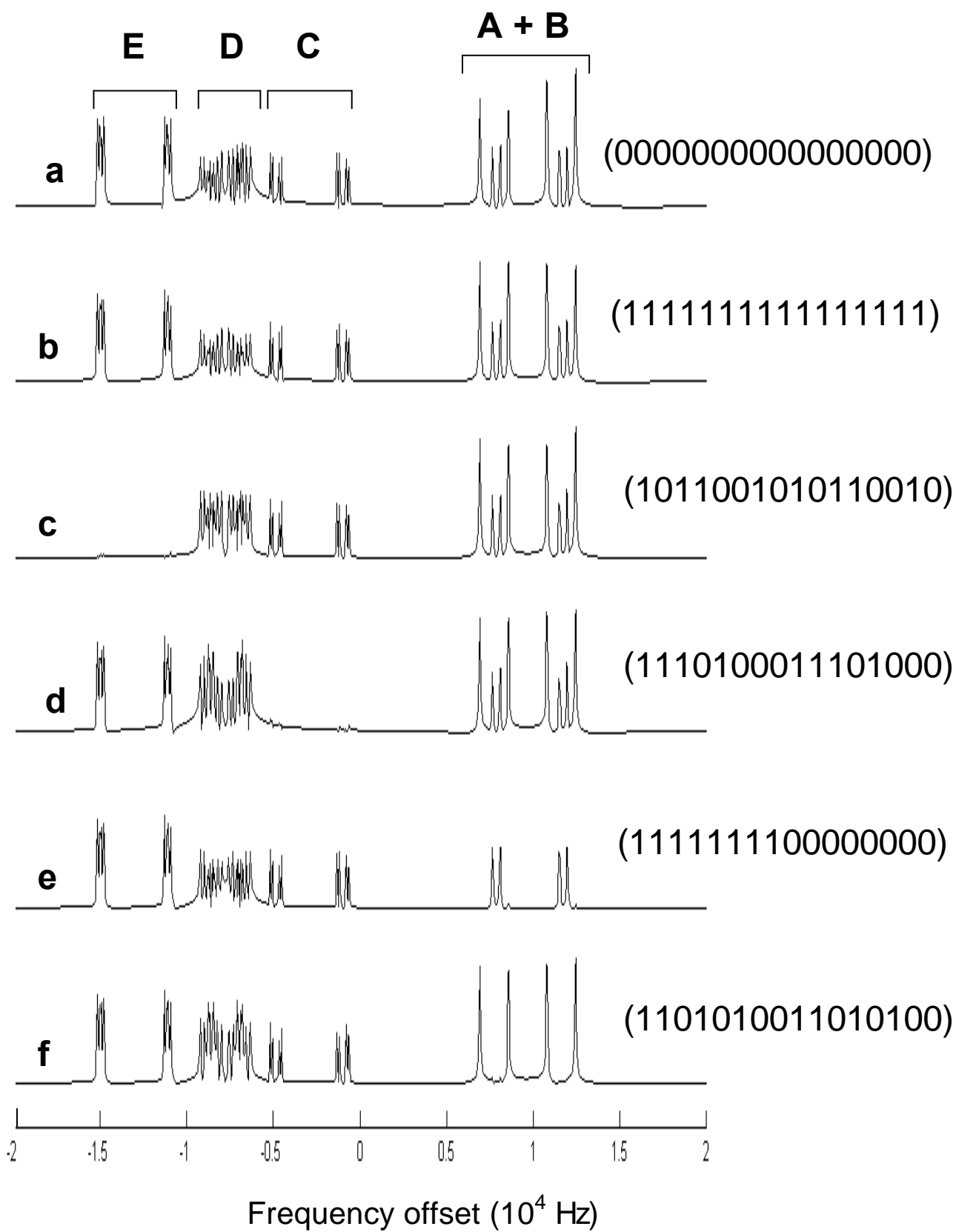

Fig. 5, V.L. Ermakov and B.M. Fung

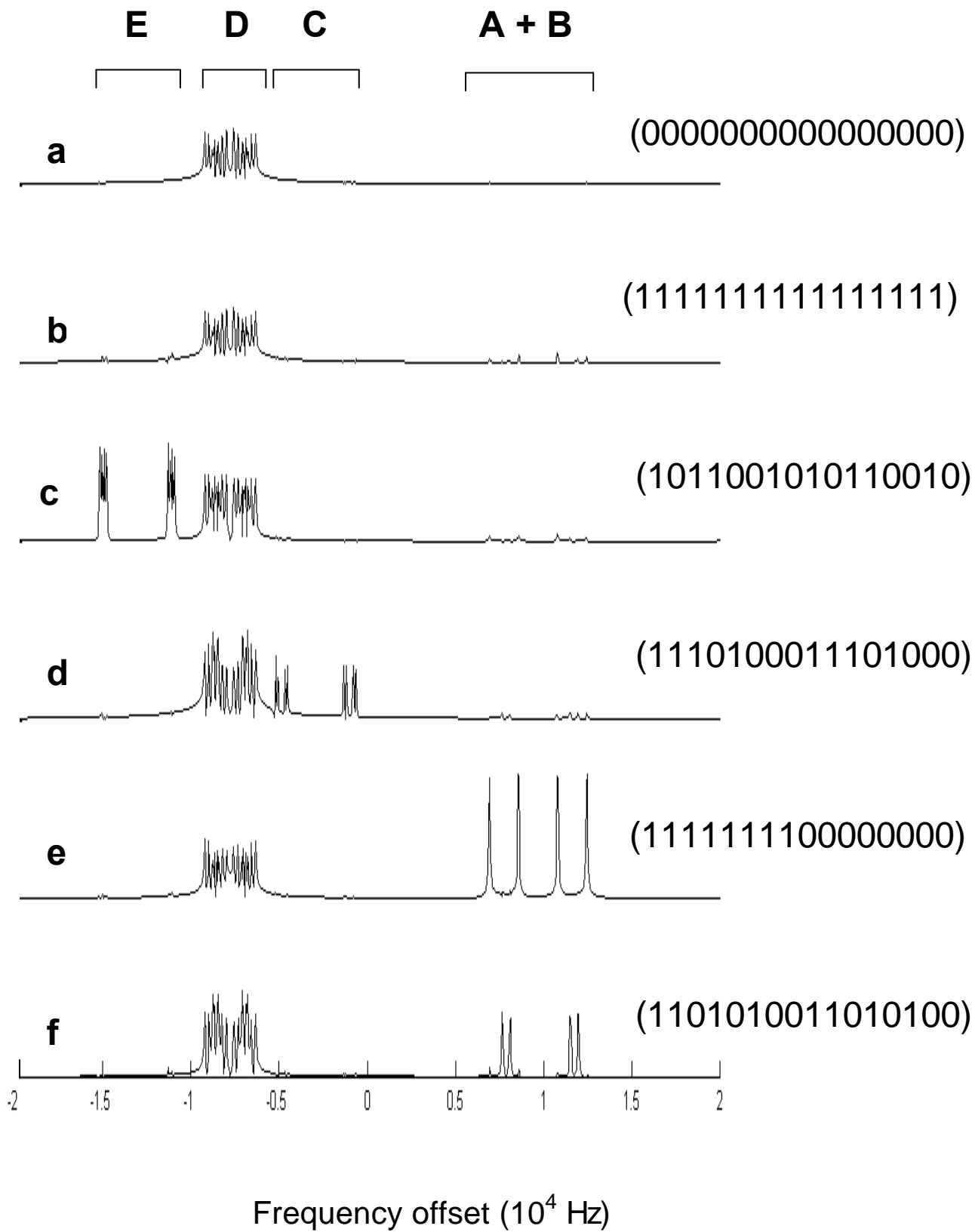

Fig. 6, V.L. Ermakov and B.M. Fung

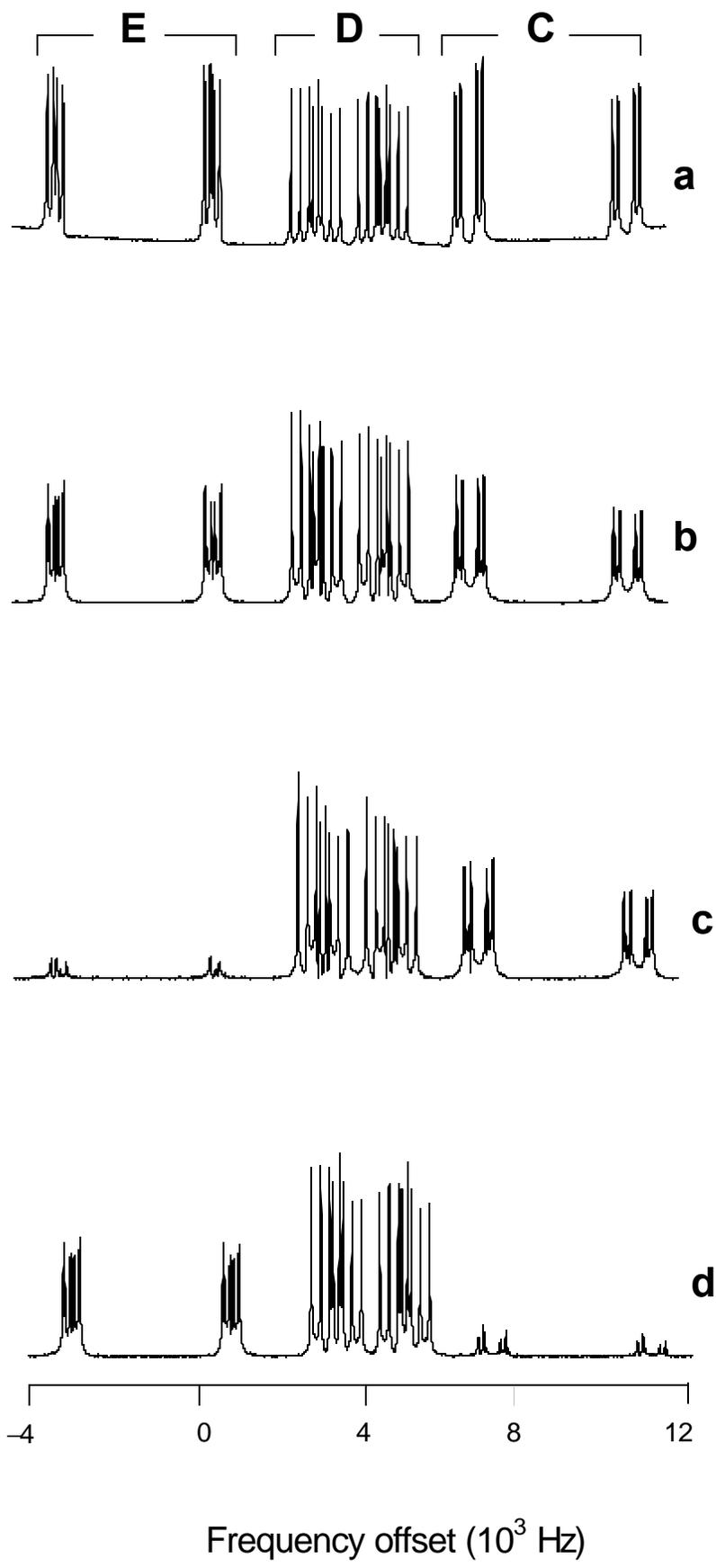

Frequency offset ($10^3$ Hz)

Fig. 7, V.L. Ermakov and B.M. Fung

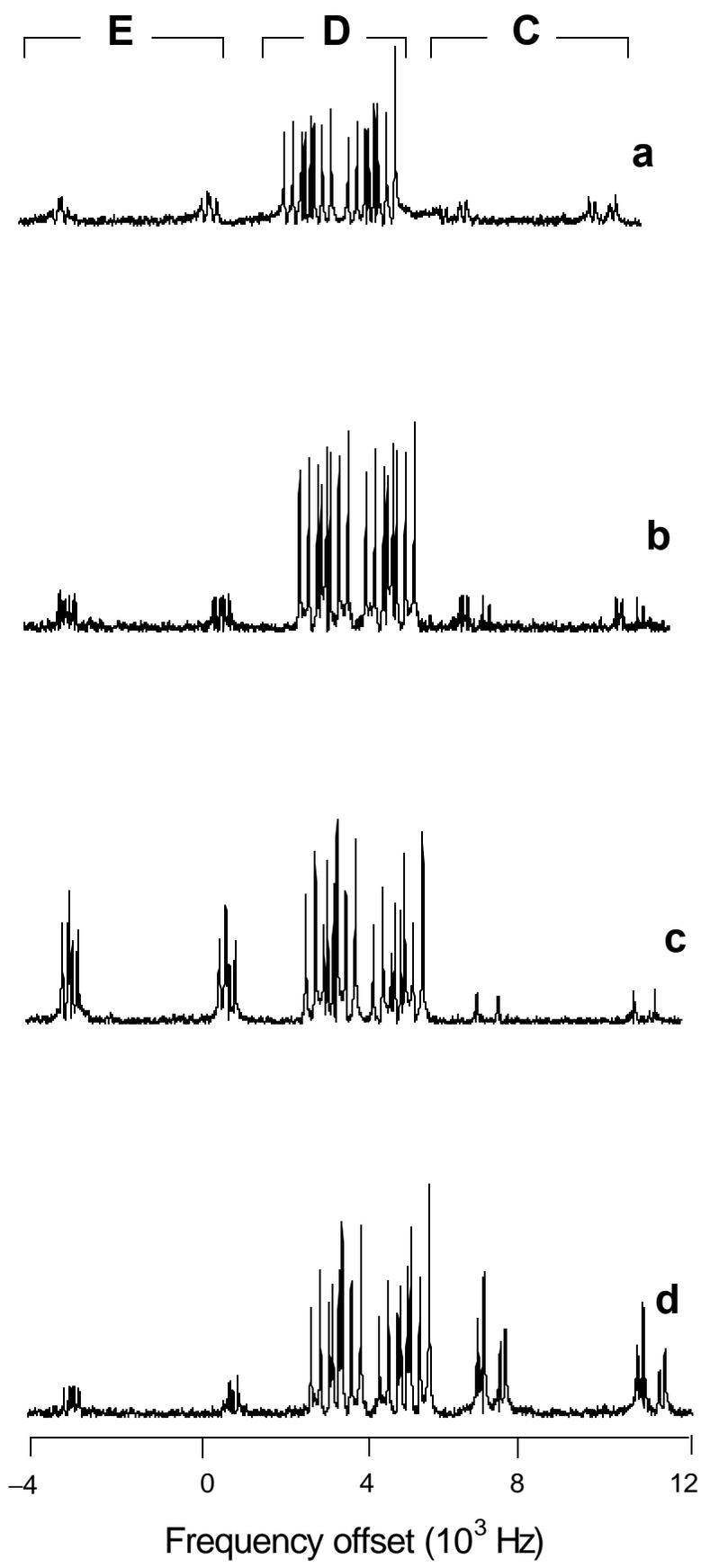

Fig. 8, V.L. Ermakov and B.M. Fung